\documentclass[]{emulateapj}
\usepackage{epsfig}

\usepackage{multirow}

\usepackage{enumitem}

\usepackage{rotating}
\usepackage{natbib}
\usepackage{latexsym}
\bibpunct{(}{)}{;}{a}{}{,}

\begin{document}

\newcommand{\ledd}{%
$L_{Edd}$}

\def\rem#1{{\bf #1}}
\def\hide#1{}

\newcommand{\specialcell}[2][c]{%
  \begin{tabular}[#1]{@{}c@{}}#2\end{tabular}}

\defcitealias{Kuulkers10}{K10}

\title{The Fermi-GBM X-ray burst monitor: thermonuclear bursts
from 4U~0614+09}

\author{M. Linares\altaffilmark{1,7}, V. Connaughton\altaffilmark{2},
P. Jenke\altaffilmark{3,8}, A.~J. van der Horst\altaffilmark{4},
A. Camero-Arranz\altaffilmark{5,9}, C. Kouveliotou\altaffilmark{3},
D. Chakrabarty\altaffilmark{1}, E. Beklen\altaffilmark{10},
P.~N. Bhat\altaffilmark{2}, M.~S. Briggs\altaffilmark{2},
M. Finger\altaffilmark{5}, W.~S. Paciesas\altaffilmark{5},
R. Preece\altaffilmark{2}, A. von Kienlin\altaffilmark{6},
C.~A. Wilson-Hodge\altaffilmark{3}}

\submitted{Accepted for publication in The Astrophysical Journal.}

\altaffiltext{1}{Massachusetts Institute of Technology - Kavli
Institute for Astrophysics and Space Research, Cambridge, MA 02139,
USA}

\altaffiltext{2}{University of Alabama in Huntsville, Huntsville, AL
35899, USA}

\altaffiltext{3}{Space Science Office, VP62, NASA/Marshall Space
Flight Center, Huntsville, AL 35812, USA}

\altaffiltext{4}{Astronomical Institute `Anton Pannekoek', University
of Amsterdam, 1090 GE Amsterdam, the Netherlands}

\altaffiltext{5}{Universities Space Research Association, Huntsville,
AL 35805, USA}

\altaffiltext{6}{Max Planck Institute for Extraterrestrial Physics,
Giessenbachstrasse, Postfach 1312, 85748 Garching, Germany}

\altaffiltext{7}{NWO Rubicon Fellow}

\altaffiltext{8}{NASA Postdoctoral Program Fellow}

\altaffiltext{9}{Current address: Institut de
Ci{\`e}ncies de l'Espai (IEEC-CSIC), Campus UAB, Fac. de Ci{\`e}ncies,
Torre C5, parell, Barcelona, 08193, Spain}

\altaffiltext{10}{Physics Department, Suleyman Demirel University, 32260
Isparta, Turkey}


\begin{abstract}

Thermonuclear bursts from slowly accreting neutron stars (NSs) have
proven difficult to detect, yet they are potential probes of the
thermal properties of the neutron star interior.
During the first year of a systematic all-sky search for X-ray bursts
using the Gamma-ray Burst Monitor (GBM) aboard the {\it Fermi
Gamma-ray Space Telescope} we have detected 15 thermonuclear bursts
from the NS low-mass X-ray binary 4U~0614+09, when it was
accreting at nearly 1\% of the Eddington limit.
We measured an average burst recurrence time of 12$\pm$3~d (68\%
confidence interval) between March 2010 and March 2011, classified all
bursts as normal duration bursts and placed a lower limit on the
recurrence time of long/intermediate bursts of 62~d (95\% confidence
level).
We discuss how observations of thermonuclear bursts in the hard X-ray
band compare to pointed soft X-ray
observations, and quantify such bandpass effects on measurements of
burst radiated energy and duration.
We put our results for 4U~0614+09 in the context of other bursters and
briefly discuss the constraints on ignition models.
Interestingly, we find that the burst {\it energies} in 4U~0614+09 are
on average between those of normal duration bursts and those measured
in long/intermediate bursts. 
Such a continuous distribution in burst energy provides a new
observational link between normal and long/intermediate bursts.
We suggest that the apparent bimodal distribution that defined normal
and long/intermediate {\it duration} bursts during the last decade
could be due to an observational bias towards detecting only the
longest and most energetic bursts from slowly accreting NSs.

\end{abstract}

\maketitle

\section{Introduction}
\label{sec:intro}

Matter accreted onto neutron stars (NSs) in low-mass X-ray binaries
(LMXBs) piles up and settles into the NS envelope. After a period
accumulating fuel, thermonuclear reactions become unstable and the
accreted layer ignites, producing a thermonuclear flash
\citep[also known as type I X-ray burst, ``burst''
hereafter;][]{Grindlay76,Maraschi77}. During the thermonuclear runaway
all or part of the accreted hydrogen (H) and/or helium (He) are
converted into heavier elements so that fresh fuel needs to accumulate
before the next burst occurs, giving rise to a cyclic behavior. This
has been observed in many thermonuclear burst sources (``bursters'')
that alternate short bursts (typically seconds- to minutes-long) with
extended periods of persistent accretion-powered emission
\citep[typically hours--days; see][for reviews]{Lewin93,Strohmayer06}.
The persistent luminosity, $L_\mathrm{pers}$, often expressed as a
fraction of the Eddington limit, is thought to trace the mass
accretion rate per unit surface area, $\dot{m}$.
When the $\dot{m}$ history is known, the time spanned between two
successive bursts (known as burst wait time,
$t_\mathrm{wait}$) determines how much mass needs to be piled up
until the ignition conditions are met at the base of the accreted
layer.
When data gaps prevent the detection of successive bursts a more
convenient parameter is the wait time averaged over the available
observations, i.e., the ratio between total accumulated exposure time
and number of bursts detected, which we refer to as burst recurrence
time, $t_\mathrm{rec}$.

The vast majority of thermonuclear bursts observed during more than
three decades, referred to as {\it normal bursts}, last between a few
seconds and a few minutes and radiate between 10$^{38}$~erg and a few
times 10$^{39}$~erg \citep[e.g.,][]{Galloway08}.
In the last decade two new classes of bursts have emerged: {\it
long/intermediate bursts} \citep[long duration bursts or simply ``long
bursts''; e.g., ][]{intZand05}, which last tens of minutes and radiate
10$^{40}$--10$^{41}$~erg; and {\it superbursts}, which last for hours
up to more than a day and radiate energies of the order of
10$^{42}$~erg \citep{Cornelisse00,Kuulkers04,Chenevez08}.
As more long bursts have been discovered, the question was raised of
whether they form a distinct population of bursts or they constitute
a rare, extremely long and energetic case of normal bursts.

Long bursts take place preferentially when $\dot{m}$ is low, near or
below 1\% of the Eddington rate, thus their properties depend on
the nuclear burning processes at work at the lowest accretion rates,
where sedimentation of heavy elements may play a role
\citep{Peng07}. Furthermore, ignition of long bursts takes place deep
in the envelope after a thick layer of fuel is built up. Hence their
properties are also sensitive to the heat flowing from the crust,
providing an independent method to study the thermal properties of the
NS interior \citep{Cumming06}.
We refer hereafter to bursts occurring when $\dot{m}$ is close to
or lower than 1\% of the Eddington limit as {\it low-$\dot{m}$
bursts}, with the understanding that these include both normal and
long duration bursts.

4U~0614+09 harbors a slowly and persistently accreting NS
\citep[active since its discovery,][]{Giacconi72} and belongs to the
sub-class of low luminosity NS-LMXBs \citep[the so-called atoll
sources,][]{HK89}. The system is likely an ultracompact binary, based
on the optical-to-X-ray flux ratio \citep{vanParadijs94b,Juett01},
tentative orbital periodicities \citep{Obrien05,Shahbaz08} and its low
but persistent X-ray luminosity \citep{intZand07}. However, the exact
nature and composition of the (likely degenerate) donor star has been
a matter of debate \citep[although strong constraints have been placed
from X-ray and optical spectroscopy; ][]{Juett01,Nelemans04,Werner06}.

Although 4U~0614+09 is a known burster \citep{Swank78,Brandt92} and a
frequently observed source, a relatively small number of bursts have
been recorded from 4U~0614+09 since its discovery four decades
ago. Only two bursts were detected with the {\it Rossi X-ray Timing
Explorer} (RXTE) despite numerous observations over the course of
16~yr (which amount to more than 2~Msec of on-source exposure time). A
compilation of these and 31 other bursts detected with several X-ray
observatories was presented by \citet[][\citetalias{Kuulkers10}
hereafter]{Kuulkers10}. They found two long duration bursts during
2001--2002 with integrated radiated energies $E_\mathrm{b} \gtrsim
$4$\times$10$^{40}$~erg, when the persistent soft X-ray luminosity was
steady, and an increased normal burst rate after that period with the
corresponding $t_\mathrm{rec}$ in the 9--14~d range.

The scarcity of bursts detected from 4U~0614+09 or similar systems is
not surprising. Low-$\dot{m}$ bursts are expected to recur on
timescales much longer than the observing windows of pointed X-ray
telescopes, which typically observe LMXBs with little or no
interruptions for up to about a day. Hence detection with pointed
instruments is rare and both $t_\mathrm{wait}$ and $t_\mathrm{rec}$
have remained poorly constrained.
Thanks to its very large instantaneous field of view ($\simeq$10~sr) and
its sensitivity to X-ray photon energies down to 8~keV, the Gamma-ray
Burst Monitor \citep[GBM;][]{Meegan09} on board the {\it Fermi
Gamma-ray Space Telescope} is able to detect rare, bright and
short-lived X-ray bursts. This is of particular relevance to the study
of thermonuclear bursts from accreting NSs.
A dedicated search using wide-field X-ray monitors provides the most
efficient way to measure accurately the recurrence time of
low-$\dot{m}$ bursts.
With this goal, we started on 2010 March 12 a systematic GBM all-sky
search for X-ray bursts, as described in Section~\ref{sec:data}.
We report 15 thermonuclear bursts from 4U~0614+09 collected during
the first year of our campaign, between 2010 March 12 and 2011 March
12.
We present the burst properties and accretion rate history in
Section~\ref{sec:results}. In Section~\ref{sec:discussion} we discuss
these results, quantify bandpass effects on the observed burst
properties (Section~\ref{sec:band}) and compare 4U~0614+09 to other
low-$\dot{m}$ bursters (Section~\ref{sec:beef}).
Section~\ref{sec:summary} summarizes our main findings and
conclusions.

\section{Data Analysis}
\label{sec:data}

\subsection{GBM all-sky search for X-ray bursts}
\label{sec:search}

GBM is an all sky monitor whose primary objective is to extend the
energy range over which gamma-ray bursts (GRBs) are observed in the
Large Area Telescope (LAT) on Fermi \citep{Meegan09}. The twelve
sodium iodide (NaI) scintillation detectors cover an energy range from
8 keV to 1 MeV. Due to the spacecraft's 50 degree rocking angle,
exposure of the sky is essentially uniform throughout the current
search period between 2010 March 12 and 2011 March 12. The X-ray burst
search uses the first 3 channels (8--50~keV) of CTIME data from the
NaI detectors. Periods where the spacecraft is performing a rapid
maneuver ($>$2.75$\times 10^{-3}$ rad/s), SAA passages, GRBs and
contaminating features, such as solar events, that prevent a good
background model fit are removed from the data.  An empirical
background model which includes Earth occultations of bright sources
is fit to the data. Short-lived ($\sim$10--1000~s) X-ray flares are
selected by visually inspecting the CTIME Channel 1 data (12--25 keV)
along with the fitted background model. Using the first three
channels, a preliminary localization is obtained using the cosine
differential response of the detectors (see Sec.~\ref{sec:loc}).
If the $\chi^2$ of the localization is less than 450, the localization
is more than 10$^\circ$ or 3$\sigma$ away from the Sun and the
position lies above and more than 1$\sigma$ from the Earth's limb (a
66$^\circ$ radius circle around the geocenter), the selected feature is
considered an X-ray burst candidate and set aside for further
analysis. We define the observing duty cycle as the ratio between the
time when the source was visible to GBM (excluding times of Earth
occultation, rapid slews and SAA passages) and the total time spanned
by the current search (one year). In the case of 4U 0614+09, the duty
cycle was 49.3\% and the total effective exposure time 180~d.

In order to optimize the signal-to-noise ratio (S/N) in the X-ray
bursts from 4U~0614+09 and measure the burst timescales and peak count
rate, we extracted background-corrected 12--25 keV light curves adding
the three brightest detectors, using daily CTIME data rebinned to 1s
time resolution. We defined the burst rise time $t_\mathrm{rise}$ as
the time for the resulting intensity to increase from 25\%
to 90\% of its peak value. We interpolated linearly between data
points and we estimate an uncertainty of 0.5s (half the time
resolution used) in $t_\mathrm{rise}$. We used CSPEC light curves in
the 8--15~keV band keeping the original 4s time resolution in order to
measure the duration of the bursts, defining the burst end as the time
when the intensity drops below 10\% of the peak.

\begin{table*}[t]
\caption{GBM X-ray bursts from 4U~0614+09.}
\begin{minipage}{\textwidth}
\begin{center}
\begin{tabular}{ c c c c c c c }
\hline\hline
ID & Peak time & Detectors\footnote{List of detectors pointing within 60$^\circ$ of 4U~0614+09 at the time of the burst, used for spectral analysis (Sec.~\ref{sec:spec}). The angular distance from each detector to 4U~0614+09 is given between parenthesis, in degrees. Only detectors with pointing offsets $\leq$50$^\circ$ were used to extract the light curves.} & Peak rate\footnote{Net 12--25~keV count rate from the three brightest detectors.} & R.A. & DEC & Error\\
 & (UTC) & n($^\circ$) & (c/s) & ($^\circ$) & ($^\circ$) & ($^\circ$)\\
\hline
B1 & 2010-03-28 00:16:13 & 9(22) 0(42) 1(44) 10(45) 2(59) & 381$\pm$37 & 98.4 & 4.6 & 3.4 \\
B2 & 2010-04-29 03:46:49 & 1(22) 2(26) 0(45) 5(59) & 617$\pm$41 & 92.6 & 8.5 & 6.0 \\
B3 & 2010-05-18 10:56:59 & 4(29) 5(32) 3(50) & 360$\pm$37 & 95.0 & 4.4 & 9.3 \\
B4 & 2010-05-25 07:35:59 & 1(14) 0(19) 3(46) 5(57) 2(59) & 595$\pm$40 & 104.6 & 1.4 & 6.8 \\
B5 & 2010-07-05 05:08:03 & 5(16) 1(43) 2(51) 3(55) 0(60) & 593$\pm$41 & 87.8 & 6.6 & 5.5\\
B6 & 2010-07-29 07:12:47 & 1(26) 2(29) 5(39) 0(50) & 496$\pm$39 & 99.0 & 10.5 & 8.2 \\
B7 & 2010-08-09 09:30:02 & 4(17) 3(36) 5(46) & 740$\pm$43 & 87.9 & 10.1 & 2.9\\
B8 & 2010-09-10 18:20:15 & 3(14) 4(33) 5(60) & 489$\pm$35 & 96.7 & 15.2 & 4.5 \\
B9 & 2010-10-12 03:54:43 & 10(37) 2(37) 9(46) 1(47) & 311$\pm$34 & 94.9 & 13.7 & 6.2 \\
B10
%
 & 2010-10-21 09:33:42 & 7(23) 6(32) 3(42) 8(46) 4(60) & 596$\pm$41 & 94.3 & 9.6 & 3.5 \\
B11 & 2010-11-28 03:50:45 & 9(27) 10(30) 11(41) & 668$\pm$40 & 97.0 & 12.0 & 4.1 \\
B12 & 2011-01-11 20:29:51 & 9(18) 11(44) 10(45) 6(51) 7(54) & 490$\pm$39 & 87.0 & 6.8 & 3.6 \\
B13 & 2011-01-14 15:26:50 & 9(27) 11(36) 10(40) 7(59) & 302$\pm$36 &  97.7 & 18.4 & 9.7 \\
B14 & 2011-01-18 22:47:38 & 7(12) 6(17) 9(50) 0(57) 8(57) 11(60) & 627$\pm$41 & 98.5 & 13.2 & 10.4 \\
B15 & 2011-01-24 11:00:23 & 9(22) 10(31) 11(46) & 716$\pm$42 & 89.8 & 9.4 & 7.8 \\
\hline\hline
\end{tabular}
\end{center}
\end{minipage}
\label{table:bursts}
\end{table*}

\subsection{Localization}
\label{sec:loc}

The 12 NaI detectors are thin disks with different orientations. Using
the known angular response of the detectors, the relative count rates
registered during a burst in the 12 detectors provide a solution for
the likely arrival direction of the burst. The measured count rates
are compared with model rates for arrival directions on a $1^\circ$
resolution grid measured in the spacecraft coordinate frame. A
$\chi^2$-minimization algorithm finds the best match for the observed
rates to 41168 positions on the sky, with an uncertainty that depends
on how steeply $\chi^2$ increases away from the minimum.
In order to make the assessment of the goodness-of-fit independent of
the intensity of the burst, the observed rates are first normalized to
a fiducial "average" burst before calculating $\chi^2$. This reduces
the $\chi^2$ values for very bright bursts with low statistical
uncertainties where the systematic errors dominate, and allows us to
use one $\chi^2$ value to reject poor fits to the data over the range
of burst intensities detected by GBM.

Because the angular response is energy-dependent, a source spectrum is
folded through the response when calculating the model rates. The
model rates in the search for X-ray bursts are calculated at energies
below 50 keV and compared to count rates measured in the corresponding
CTIME channels. To maximize the signal in the search for these events,
the counts in CTIME channels 0 through 2 are used, and the model rates
are evaluated between 8 and 50 keV. The uncertainties in the detector
responses below 10 keV and the noisier background at the lowest
energies may introduce larger systematic errors than incurred above 10
keV, so the localization is also performed in the 10--50 keV energy
band. By default the search for X-ray bursts uses a power-law model
rate function with an index of 2, which represents a variety of
phenomena but may not be the best model for type I X-ray bursts.
Model rates for a softer power-law function with index 3 were
therefore used at the localization stage once the X-ray burst
candidates were isolated in the overall data set. Model rates with a
blackbody spectrum that was later shown to be a good fit to the X-ray
burst candidates were subsequently generated and used to see whether a
good model fit produced a better localization.

As expected, the statistical error on the localization was higher by
about $1^\circ$ for a given model when the rates below 10 keV were
excluded from the localization of the 15 events associated with
4U~0614+09. The error on the localization, inferred from the distance
to the known source position, was comparable for localizations
performed using both energy ranges. The ratio between the offset from
the source and the statistical uncertainty suggests that the
systematic error on the localizations using the 8--50~keV energy range
may be larger, with 41\% (82\%) contained within the $1 (2) \sigma$
uncertainty region compared to 54\% (87\%) for the localizations
performed in the 10--50~keV energy range.
The centroids (weighting each position by its statistical uncertainty)
of the 15 combined locations are between 0.5$^\circ$--1.1$^\circ$ from
4U~0614+09 for the 10--50~keV range localizations depending on the
model, with the more representative softer power-law and blackbody
spectral models yielding centroids at 0.5 and $0.6^\circ$ from the
source. The blackbody and softer power-law models produced
localizations that were on average closer to the source by about
$0.7^\circ$ compared to the power-law model with index of -2. The
8--50~keV localizations produce a centroid that lies between $1.6$ and
$1.9^\circ$ from the source.
Within the relatively low (15) number of events available, our results
suggest that systematic errors may be larger when the full energy
range is used to localize events. Whilst our analysis suggests that
using the 10--50~keV energy range with either a soft power-law or
black-mody model produces the best results when localizing type I
X-ray bursts using GBM data, the events are weak enough that the
uncertainties are dominated by statistics, and these systematic
effects are small by comparison.

\subsection{Spectral analysis}
\label{sec:spec}

We performed time-averaged and time-resolved spectroscopy of the 15
confirmed bursts from 4U~0614+09. We used the continuous CSPEC data
type in our analysis, with the best available spectral resolution for
GBM and 4~s time resolution \citep[][]{Meegan09}. The data were
analyzed with the RMFIT (3.3rc8) software package which has been
developed for the GBM data analysis. The time-varying
background was fitted with polynomial functions of the third or fourth
order. We generated Detector Response Matrices using GBMRSP V1.9. In
our analysis we used all the NaI detectors with zenith angles to
the source smaller than $60^{\circ}$, and which view was not
obstructed by the solar panels or other parts of the spacecraft.

We have fitted the time-integrated spectra with a power law and a
blackbody function, using the Castor statistic
(C-stat)\footnote{A modification of Cash's C-statistic used to
fit spectra with low number of counts, which tends to $\chi^2$ for
large number of counts;
http://heasarc.gsfc.nasa.gov/docs/xanadu/xspec/wstat.ps}, and in all
bursts a blackbody provided the best fit to the data. The difference
in C-stat between the two models ranges from several tens to hundreds,
illustrating the improvement of a thermal model over a power
law. 
To quantify this and estimate the significance of the blackbody
fit improvement over the power law fit, we simulated 10$^4$ spectra
from the best-fit model parameters. We found that for a typical value
($\simeq$40) of the difference in C-stat between thermal and power law
model fits, the fit improvement is significant at the $>$99.99\%
(3.9$\sigma$) confidence level.
Since these events are spectrally very soft for the GBM energy
band, we used the 8--30~keV range in our analysis, but we note
that using the full NaI energy range gives consistent results. We also
checked spectral fits from individual detectors and they all give
consistent results. Here we use the results obtained by combining the
appropriate detectors to maximize the S/N.

\subsection{Persistent emission and mass accretion rate}
\label{sec:pers}

In order to characterize the soft and hard persistent
(accretion-powered) X-ray emission from 4U~0614+09 and estimate
$\dot{m}$, we used 1.5--12~keV light curves collected by the All-Sky
Monitor \citep[ASM;][]{Levine96} on board the {\it Rossi X-ray Timing
Explorer} as well as 15--50~keV light curves from {\it Swift}'s Burst
Alert Telescope \citep[BAT;][]{Barthelmy05}, using data from the hard
X-ray transient monitor \citep{Krimm06}. We used ASM and BAT daily
intensity measurements taken between 2010 March 12 and 2011 March 12
including all available (2-$\sigma$) detections, which amount to 81\%
and 76\% of the whole one-year period for ASM and BAT, respectively.

As we are interested for the purpose of this work in long-term (weeks
to months) changes in $\dot{m}$ as well as its mean value over the
search period, we applied a running average with a 12-d window to the
resulting light curves. At the expense of disregarding variability on
timescales of less than $\simeq$2 weeks, this method increases the S/N
and removes short data gaps due to, e.g., high background in ASM or
BAT. The window length was choosen to prevent data gaps from affecting
the running average, while keeping a relatively short window. We then
converted the ASM and BAT intensity into 2--20~keV and 20--100~keV
luminosity, respectively, using WebPimms\footnote{Portable Interactive
Multi-Mission Simulator v. 4.5 at
http://heasarc.gsfc.nasa.gov/Tools/w3pimms.html} and a distance to
4U~0614+09 of 3.2~kpc \citepalias{Kuulkers10}. We assumed that the
X-ray spectrum of 4U~0614+09, averaged over its spectral states
\citep{Linares09d}, is well represented by a power law model with
photon index 2. The resulting 2--100~keV luminosity gives an estimate
of the bolometric persistent luminosity, $L_\mathrm{pers}$, which we
convert into $\dot{m}$ assuming that all the gravitational energy is
radiated away isotropically at the surface of a 10~km radius,
1.4~$M_\odot$ mass NS \citep[this and other assumptions may
result into uncertainties in $\dot{m}$ of a factor $\sim$2; see,
e.g., discussion in][]{Coriat12}.
When normalizing $L_\mathrm{pers}$ by the Eddington luminosity
($L_\mathrm{Edd}$) we use a fiducial value of
2.5$\times$10$^{38}$~erg~s$^{-1}$, yet we note that the exact value
depends on the composition of the accreted material. Using
3.8$\times$10$^{38}$~erg~s$^{-1}$, a value more typical of H-poor
material, does not change our results and conclusions.

\begin{figure}[ht!]
\centering
  \resizebox{1.1\columnwidth}{!}{\rotatebox{0}{\includegraphics[]{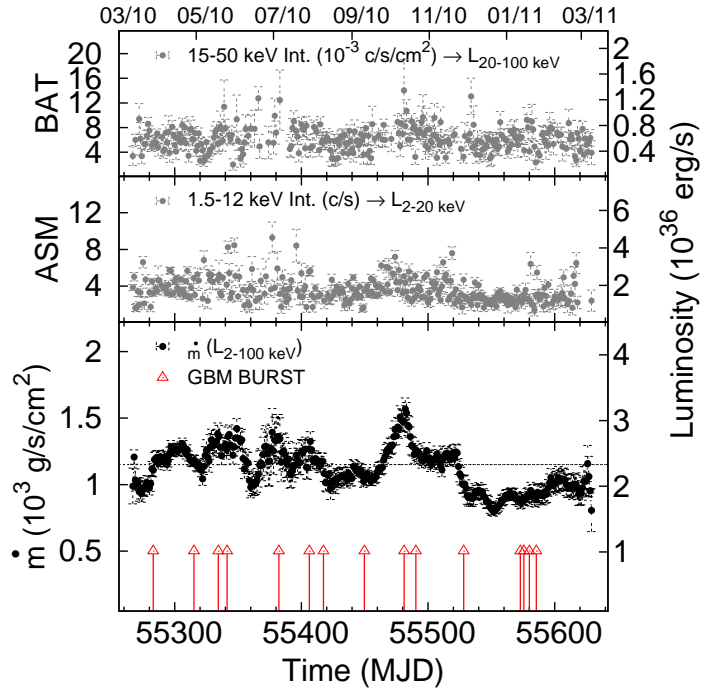}}}
  \caption{
{\it Bottom panel:} Evolution of mass accretion rate per unit surface
  (left axis) derived from the 2--100~keV persistent luminosity (right
  axis) after applying a 12-d window running average (see
  Sec.~\ref{sec:pers} for details). Red arrows along bottom axis show
  the times of the 15 thermonuclear bursts from 4U~0614+09 detected
  with GBM during the 1~yr period. Month start dates are marked along
  the top axis, from March 2010 to March 2011.
{\it Top \& middle panels:} Hard (BAT) and soft (ASM) X-ray light
  curves of 4U~0614+09. Daily-averaged intensity measurements (left
  axes) are converted into luminosity (right axes).
}
    \label{fig:lc}
\end{figure}

\section{Results}
\label{sec:results}

\begin{table}[t]
\scriptsize
\caption{Properties of the GBM type I X-ray bursts from 4U~0614+09.}
\begin{center}
\begin{tabular}{ c c c c c c c c}
\hline\hline
\footnotetext{Burst peak luminosity $L_\mathrm{peak}$ and radiated
energy $E$ at 3.2~kpc. Rise time, duration
%
and $E$ as measured by GBM, not corrected for
bandpass. $\Delta t$ is the time in days since the previous burst from
4U~0614+09 detected by GBM (not corrected by on-source exposure time
and therefore corresponds to an upper limit on the burst wait
time, $t_\mathrm{wait}$; Sec.~\ref{sec:intro}).}
ID & kT$_\mathrm{peak}$ & L$_\mathrm{peak}$ & E & Rise & Duration & $\Delta t$\\
 & (keV) & (10$^{38}$ erg/s) & (10$^{39}$~erg) & (s) & (s) & (d)\\
\hline
B1 & 2.7$\pm$0.1 & 1.79$\pm$0.07 & 2.07$\pm$0.01 & 1.6 & 28.6 & --\\
B2 & 3.0$\pm$0.1 & 1.65$\pm$0.05 & 1.77$\pm$0.02 & 2.2 & 20.3 & 32.1\\
B3 & 2.9$\pm$0.2 & 0.89$\pm$0.07 & 0.72$\pm$0.03 & 0.9 & 29.4 & 19.3\\
B4 & 2.9$\pm$0.1 & 1.75$\pm$0.05 & 1.06$\pm$0.03 & 1.2 & 11.1 & 26.2\\
B5 & 2.6$\pm$0.1 & 1.97$\pm$0.07 & 2.33$\pm$0.12 & 1.5 & 26.0 & 40.9\\
B6 & 3.1$\pm$0.1 & 1.28$\pm$0.06 & 0.82$\pm$0.02 & 1.1 & 16.4 & 24.4\\
B7 & 3.2$\pm$0.1 & 2.72$\pm$0.06 & 5.77$\pm$0.03 & 4.5 & 46.5 & 35.2\\
B8 & 3.0$\pm$0.1 & 1.91$\pm$0.05 & 4.19$\pm$0.03 & 1.8 & 44.0 & 32.4\\
B9 & 2.7$\pm$0.1 & 1.38$\pm$0.06 & 2.23$\pm$0.02 & 1.3 & 34.1 & 31.4\\
B10 & 3.0$\pm$0.1 & 2.36$\pm$0.09 & 2.15$\pm$0.06 & 1.5 & 20.7 & 9.2\\
B11 & 3.2$\pm$0.1 & 2.17$\pm$0.07 & 2.84$\pm$0.03 & 2.8 & 30.5 & 37.8\\
B12& 3.0$\pm$0.1 & 1.89$\pm$0.08 & 2.58$\pm$0.05 & 0.9 & 49.0 & 44.7\\
B13\footnote{Due to the low number of counts per spectrum, blackbody
temperature and radius are strongly correlated in B13. If we fix the
radius to a value of 5~km, consistent with all other bursts, the peak
temperature becomes kT$_\mathrm{peak} \simeq$2.5~keV.}
& 3.3$\pm$0.2 & 0.74$\pm$0.05 & 0.43$\pm$0.02 & 1.3 & 10.1 & 2.8\\
B14& 2.9$\pm$0.1 & 2.05$\pm$0.08 & 1.49$\pm$0.07 & 2.9 & 14.0 & 4.3\\
B15& 3.2$\pm$0.1 & 2.30$\pm$0.08 & 1.82$\pm$0.08 & 3.1 & 15.0 & 5.5\\
\hline\hline
\end{tabular}
\end{center}
\label{table:bprop}
\end{table}

\begin{figure}[h!]
\centering
  \resizebox{1.0\columnwidth}{!}{\rotatebox{-90}{\includegraphics[]{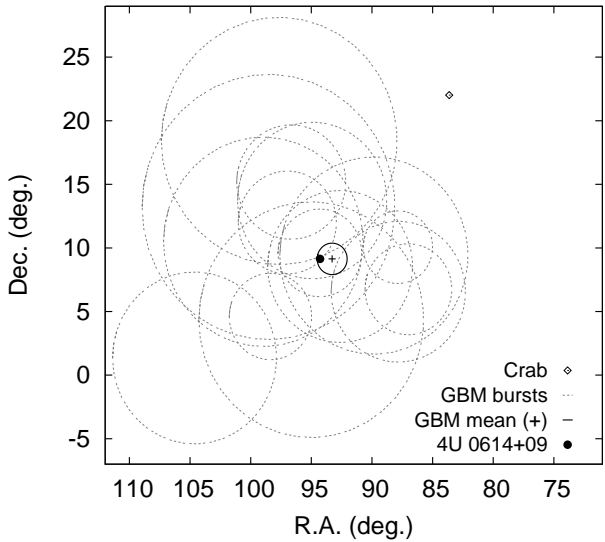}}}
  \caption{GBM localizations of the 15 bursts from 4U~0614+09
  presented in this work (gray dashed circles show 1$\sigma$ error
  circles; see Section~\ref{sec:loc} and
  Table~\ref{table:bursts}). The average position from all GBM bursts
  (black open circle), the known position of 4U~0614+09 \citep[black
  filled circle;][]{Migliari06b} and the closest bright X-ray source
  (Crab, black diamond) are also shown.}
    \label{fig:locs}
\end{figure}

We found a total of 15 X-ray bursts from 4U~0614+09 during the first
year of our {\it Fermi}-GBM X-ray burst search
(Sec.~\ref{sec:search}). Only 33 bursts had been observed from
4U~0614+09 since it was discovered, using observations taken over 32
yr by 8 different X-ray observatories \citepalias[between 1975 and
2007; see][]{Kuulkers10}. Therefore, the 15 GBM bursts presented
herein constitute a substantial improvement in burst detection
rate.
Table~\ref{table:bursts} shows all GBM bursts labeled in chronological
order B1--B15, their peak times, count rates and locations. All bursts
were clearly detected in the 12--25~keV band with all GBM NaI
detectors pointing within 60$^\circ$ of 4U~0614+09, and none of them
was detected above 50~keV, as expected for thermonuclear bursts given
their thermal spectrum. 
Figure~\ref{fig:lc} shows the burst times as well as the soft (ASM)
and hard (BAT) X-ray light curves of 4U~0614+09. From these, we
obtained the 2--100~keV persistent luminosity and inferred mass
accretion rate per unit area $\dot{m}$ (Sec.~\ref{sec:pers}), which
remained between 0.6\% and 1.3\% of the Eddington limit
($L_\mathrm{Edd}$ and $\dot{m}_\mathrm{Edd}$, respectively;
Sec.~\ref{sec:pers}) during the whole period studied in this work,
with an average value $< L_\mathrm{pers} >$=
2.3$\times$10$^{36}$~erg~s$^{-1}$ ($\dot{m}/\dot{m}_\mathrm{Edd}$ =
0.9\%). All the {\it RXTE} observations taken between March 2010 and
March 2011 showed 4U~0614+09 in the intermediate state, where it
remains most of the time \citep[the accretion state intermediate
between spectrally hard and soft states; see, e.g., ][]{Linares09d}.

As shown in Figure~\ref{fig:locs}, the burst locations are consistent
with the known position of 4U~0614+09. For about 90\% of the
bursts the 2$\sigma$ confidence region around the best location
includes 4U~0614+09 (see full details in Section~\ref{sec:loc}).
All bursts are more than 10$^\circ$ and 70$^\circ$ away from the
center of the Sun and Earth, respectively, as seen from the {\it
Fermi} spacecraft (Sec.~\ref{sec:search}). The closest known bright
X-ray source (Crab) and burster (4U~0513-40) are about 16$^\circ$ and
51$^\circ$ away, respectively, and there is no known LMXB within
15$^\circ$ of 4U~0614+09. This, together with the thermonuclear nature
of the bursts (see below), allows us to identify all bursts
B1--B15 as coming from 4U~0614+09.

\begin{figure}[ht]
\centering
  \resizebox{1.0\columnwidth}{!}{\rotatebox{0}{\includegraphics[]{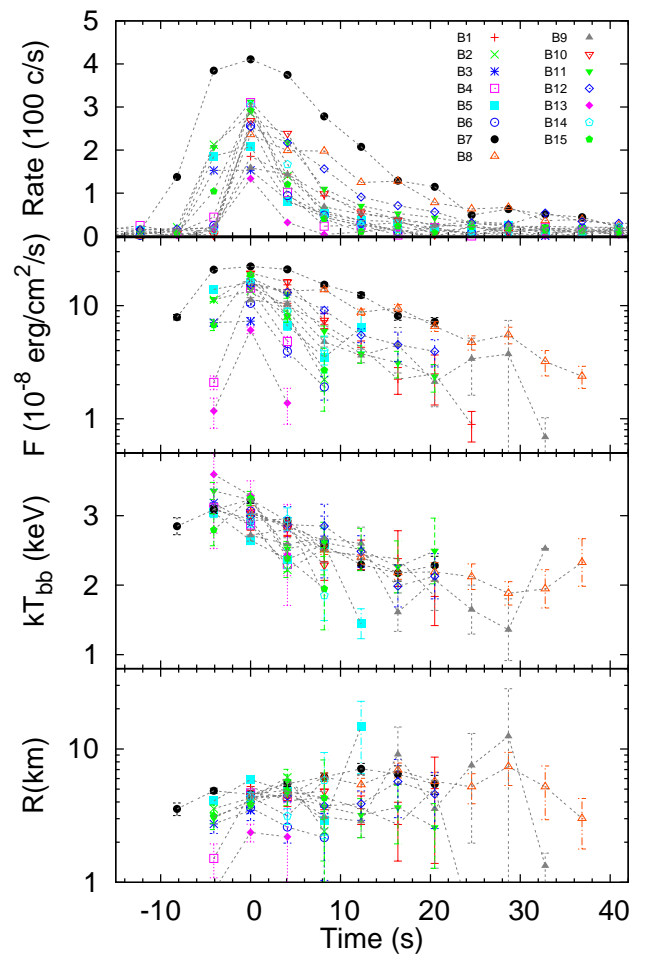}}}
  \caption{Time-resolved spectroscopy of the GBM X-ray bursts from 4U~0614+09, showing (from top to bottom): net count rate, bolometric flux, blackbody temperature and blackbody radius. The temperature decay is clearly visible, identifying all bursts as thermonuclear.}
    \label{fig:bspec}
\end{figure}

\begin{figure*}[ht!]
\centering
  \resizebox{1.4\columnwidth}{!}{\rotatebox{-90}{\includegraphics[]{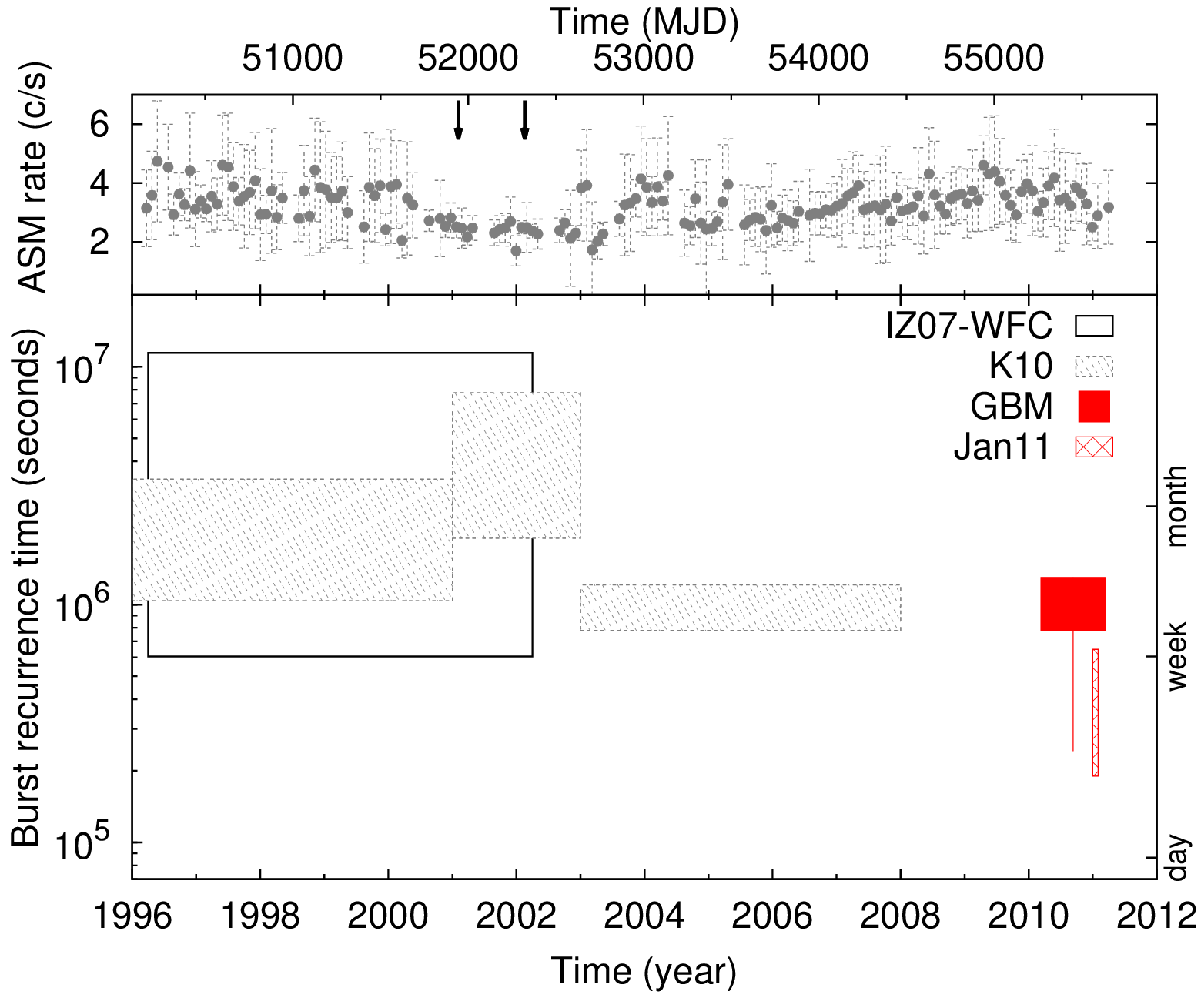}}}
  \caption{{\it Bottom:} Measurements of burst recurrence time
  $t_\mathrm{rec}$ in 4U~0614+09, between 1996 and 2012. The red
  filled rectangle shows our measurement of $t_\mathrm{rec}$ from the
  first year of the GBM all-sky X-ray burst search. The red vertical
  error bar marks the shortest wait time, 2.8~d, and the crossed red
  rectangle shows the $t_\mathrm{rec}$ measured in January 2011. Gray
  dashed rectangles display the limits on $t_\mathrm{rec}$ found by
  \citetalias{Kuulkers10}. The solid open black rectangle shows the
  WFC limits on $t_\mathrm{rec}$ for 4U~0614+09
  \citep[from][]{intZand07}. All $t_\mathrm{rec}$ ranges shown
  with rectangles are 1$\sigma$ confidence intervals assuming a
  Poisson distribution.
  {\it Top:} Mission-long {\it RXTE}-ASM light curve of 4U~0614+09,
  showing monthly weighted averages of the 1.5--12~keV
  intensity. Error bars display $\pm$1 standard deviation,
  highlighting the lower long-term variability 2001-2002 period
  discussed by \citetalias{Kuulkers10}, when the two long bursts
  were detected (their times are marked with arrows along the top
  axis).
}
    \label{fig:trectime}
\end{figure*}

We first performed power law fits to the time-averaged 8--30~keV
spectrum of all bursts and obtained photon indices in the 3--4.4
range. Such high values of the photon index reveal a relatively steep
or ``soft'' spectrum, indicative of thermal emission. Indeed, a
blackbody model gives a better fit (Sec.~\ref{sec:spec}) to the
average spectrum of all bursts, yielding temperatures in the
2.5--2.8~keV range, fully consistent with a thermonuclear origin.
The results of the time-resolved spectroscopy (Sec.~\ref{sec:spec})
are shown in Figure~\ref{fig:bspec}, and confirm the thermonuclear
nature of the bursts. We detect in all cases a significant drop in
burst temperature along the decay, also referred to as ``cooling
tail''. This is a defining property of type I X-ray bursts and a
sufficient condition to identify bursts as thermonuclear.
Burst peak temperatures ($kT_\mathrm{peak}$) are in the range
2.6--3.3~keV (Table~\ref{table:bprop}), and decay to
$\simeq$2~keV in the latest detected stages of the bursts.

We use throughout this work the 3.2~kpc distance found by
\citetalias{Kuulkers10} from the brightest photospheric radius
expansion (PRE) burst in their sample.
The apparent blackbody radii that we find (Figure~\ref{fig:bspec};
between $\simeq$3~km and $\simeq$10~km during most phases of the
bursts) are consistent with previously reported values for 4U~0614+09
\citepalias{Kuulkers10} and compatible with the expected NS emitting
area \citep[note that no gravitational redshift or spectral
corrections have been applied to these values; see,
e.g.,][]{Lewin93}. All the bursts we detect feature single-peaked
lightcurves, also when inspecting 100--200~s around the peak with
1--4~s time resolution, and we find no signs of PRE in the GBM time
resolved spectroscopy of any of the bursts (but see
Sec.~\ref{sec:band}).
The burst peak luminosities ($L_\mathrm{peak}$) that we find range
from 0.7$\times 10^{38}$~erg~s$^{-1}$ to 2.7$\times
10^{38}$~erg~s$^{-1}$ (Table~\ref{table:bprop}). The rise times we
measure in the GBM band ($t_{rise}$; Sec.~\ref{sec:search} for
definition) are between 1~s and 5~s. By integrating the
bolometric luminosity over the fraction of the bursts detected by GBM,
we find burst energies in the range [0.4--5.8]$\times
10^{39}$~erg. The burst duration in the GBM band was between
10~s and 50~s. Due to the combined effects of the GBM
bandpass and background, these are lower limits to the total radiated
burst energy and duration (see Section~\ref{sec:band} for further
discussion of bandpass corrections).

From the 15 bursts detected and the $\simeq$50\% observing duty cycle
(Sec.~\ref{sec:search}), we find a burst recurrence time between March
2010 and March 2011 of $t_\mathrm{rec}$=12$\pm$3~d (1$\sigma$
confidence interval from Poisson statistics). Notably, bursts B12 and
B13 were only 2.8~d apart, the closest burst pair ever detected from
4U~0614+09 \citepalias[shortest $t_\mathrm{wait}$ known to date was
one week;][]{Kuulkers10}. Four bursts were detected in January 2011
corresponding to a recurrence time of $\simeq$4~d, shorter than the
recurrence time during the preceding 10 months ($t_\mathrm{rec}$ was
$\simeq$13~d between 2010 March 12 and December 31). This can be seen
in Figure~\ref{fig:lc}, which shows the times of the bursts detected
by GBM together with the $\dot{m}$ evolution. Assuming a Poisson
distribution, the probability of detecting at least 4 bursts during
one month (effectively two weeks) for the average burst detection rate
1.3~month$^{-1}$ (effectively 1.3~per 2~week-period) is
3.8\%. This suggests that the intrinsic thermonuclear burst rate from
4U~0614+09 increased during January 2011. The source state during
January 2011 was similar to the rest of the search period. The
inferred $\dot{m}$ reached a minimum near January 2011 (at
$\dot{m}/\dot{m}_\mathrm{Edd}$ $\simeq$ 0.6\%).

\begin{figure*}[ht]
\centering
  \resizebox{1.6\columnwidth}{!}{\rotatebox{0}{\includegraphics[]{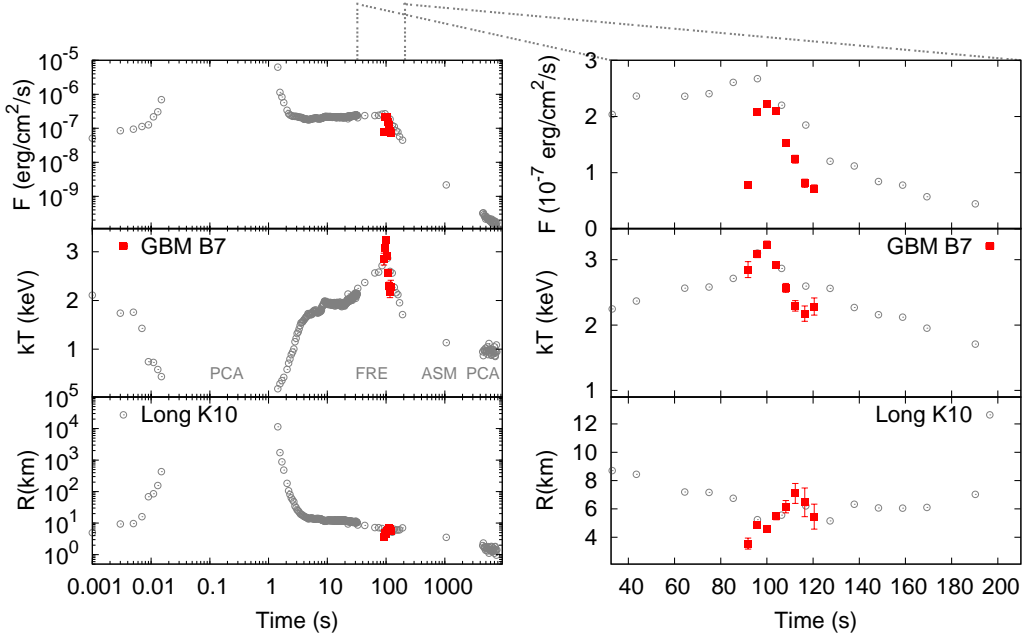}}}
  \caption{Spectral evolution during a long PRE burst from 4U~0614+09
  detected in 2001 \citepalias[open gray circles; from][burst FR
  51944]{Kuulkers10}. Panels show, {\it from top to bottom:}
  bolometric flux, blackbody temperature and radius (for a distance of
  3.2~kpc and without color or gravitational redshift corrections). We
  compare this to the spectral evolution measured by GBM during burst
  B7, detected in 2010 (filled red squares; see
  Tables~\ref{table:bursts} \& \ref{table:bprop}). {\it Left} panels
  display the entire (about 8000~s long) long burst time range. {\it
  Right} panels zoom into the touchdown phase of the burst, the only
  part hot enough to be detected in the hard X-ray band (i.e., by GBM
  and FREGATE).}
    \label{fig:iceberg}
\end{figure*}

We do not detect long/intermediate bursts from 4U~0614+09 during the
1~yr search period, and we can place a 95\% lower limit
\citep[assuming a Poisson distribution,][]{Gehrels86} on the long
burst recurrence time of 62~d.
Our measurement of $t_\mathrm{rec}$ is consistent with previous
estimates, which placed $t_\mathrm{rec}$ between one week and three
months \citep{intZand07,Kuulkers10}. Figure~\ref{fig:trectime}
compares $t_\mathrm{rec}$ from 4U~0614+09 as measured by several
instruments during the last sixteen years, between 1996 and 2012. In
only one year, the GBM results presented herein have provided the most
accurate measurement of $t_\mathrm{rec}$ and revealed the shortest
$t_\mathrm{wait}$ in 4U~0614+09. Furthermore, we are beginning to
probe changes in $t_\mathrm{rec}$ on timescales of months, as
illustrated by the likely increase in burst rate during January 2011.

\section{Discussion}
\label{sec:discussion}

We have presented the detection and properties of 15 thermonuclear
bursts from 4U~0614+09 observed between March 2010 and March 2011
within our GBM all-sky search for X-ray bursts. Using only one year of
GBM data we increased by 40\% the sample of bursts from
4U~0614+09 collected over more than three decades
\citepalias{Kuulkers10}. These results illustrate the benefits of an
X-ray monitor with instantaneous coverage of most ($\simeq$75\%) of the
sky and yield a robust measurement of the recurrence time of
low-$\dot{m}$ bursts in 4U~0614+09: $t_\mathrm{rec}$=12$\pm$3~d.
Figure~\ref{fig:trectime} shows the full 1996--2012 ASM light curve
(top panel; monthly-averaged intensity with error bars
showing its standard deviation during each month) together with
$t_\mathrm{rec}$ in 4U~0614+09. The average burst rate has been steady
over 16~yr, with two marginal changes: a decrease in burst rate in
2001--2002 reported by \citetalias{Kuulkers10}, associated with lower
variability in the ASM light curve, and the increased burst rate in
January 2011 reported herein.
The two only known long/intermediate bursts from 4U~0614+09 happened
in the 2001--2002 period, when the variability on long (days--months)
timescales was at a minimum (Fig.~\ref{fig:trectime}). This ``calm''
period has not repeated ever since, which could explain why we do not
detect long/intermediate bursts during 2010--2011.

\begin{table*}[t]
\caption{Bandpass corrections to burst energy and duration.}
\begin{minipage}{\textwidth}
\begin{center}
\begin{tabular}{ c c c c c c c c c c c}
\hline\hline
\footnotetext{Burst energy and duration for a long and normal burst
measured with three instruments in different energy bands, as
indicated. PCA and FREGATE data are taken from
\citetalias{Kuulkers10}. GBM values are obtained from simulated light
curves (Sec.~\ref{sec:band}). PCA values are considered bolometric
measurements of the burst energy and duration. Bandpass correction
factors are given as ratios between the values measured in:
FREGATE/GBM and PCA/GBM. Based on the last two columns we
estimate the bolometric corrections on burst energy and duration as
$E_\mathrm{b}/E_\mathrm{obs}$=[1.3--1.7] and
$t_\mathrm{obs}/t_\mathrm{obs}$=[1.5--55], respectively
(Sec.~\ref{sec:band}).}
%
 & \multicolumn{2}{c}{GBM} & \multicolumn{2}{c}{FREGATE} & \multicolumn{2}{c}{PCA} & \multicolumn{2}{c}{FREGATE/GBM} & \multicolumn{2}{c}{PCA/GBM}\\
 & \multicolumn{2}{c}{(8--1000~keV)} & \multicolumn{2}{c}{(6--400~keV)} & \multicolumn{2}{c}{(2--60~keV)} & & & & \\
Burst & $E_\mathrm{obs}$ (erg) & $t_\mathrm{obs}$ (s) & $E_\mathrm{obs}$ (erg) & $t_\mathrm{obs}$ (s) & $E_\mathrm{b}$ (erg) & $t_\mathrm{b}$ (s) & $E_\mathrm{obs}/E_\mathrm{obs}$ & $t_\mathrm{obs}/t_\mathrm{obs}$ & $E_\mathrm{b}/E_\mathrm{obs}$ & $t_\mathrm{b}/t_\mathrm{obs}$ \\
\hline
Normal (FR 52951) & 3.6$\times$10$^{39}$ & 30 & 6.0$\times$10$^{39}$ & 45 & ? & ? & 1.7 & 1.5 & $>$1.7 & $>$1.5 \\
Long (FR 51944) & 3.4$\times$10$^{40}$ & 140 & 3.9$\times$10$^{40}$ & 308 & 4.4$\times$10$^{40}$ & 7700 & 1.1 & 2.2 & 1.3 & 55 \\
\hline\hline
\end{tabular}
\end{center}
\end{minipage}
\label{table:bolo}
\end{table*}

\subsection{Bandpass corrections: total burst energy and duration}
\label{sec:band}

The energy spectrum of thermonuclear bursts is for most purposes well
described by a simple absorbed blackbody model.
At the peak (near the flux maximum) bursts can show a wide variety of
light curve shapes and spectral evolution. In particular, the peak of
photospheric radius expansion (PRE) bursts features a drastic increase
of the blackbody radius $R_\mathrm{bb}$ and an associated drop of the
blackbody temperature $kT_\mathrm{bb}$, due to the effect of radiation
forces on the NS photosphere \citep[e.g.,][and references
therein]{Lewin93}. Subsequently the photosphere contracts and heats up
until it reaches the NS surface again, at ``touchdown''.
Most low-$\dot{m}$ bursts are PRE events (of either normal or
long duration class) as the high ignition depths power energetic and
luminous events that can overcome the gravitational force at the
photosphere.

Soft X-ray sensitivity is necessary in order to observe the full
radius expansion phase of PRE bursts, due to the simultaneous drop in
temperature which shifts the spectrum to lower energies. Conversely,
the lack of spectroscopic evidence for PRE when observing a given
burst in the hard X-ray band does not rule out the PRE nature of the
burst. This is shown in Figure~\ref{fig:iceberg}, where we compare the
spectral evolution of a long PRE burst from 4U~0614+09 observed
in 2001 with broadband spectral coverage \citepalias[with the
PCA, ASM and FREGATE instruments; from][]{Kuulkers10} to one of the
GBM bursts presented herein. In all cases except B3 and B13 the
fluxes, temperatures and blackbody radii measured during our GBM
bursts are consistent with those measured near the ``touchdown'' phase
of a PRE burst (Figure~\ref{fig:iceberg} shows one example, B7).
Thus most thermonuclear bursts detected with GBM (and perhaps
with any hard X-ray wide field detector) are in fact consistent
with PRE bursts in the touchdown phase. However, as our GBM
X-ray burst search includes faint and untriggered events we are
also sensitive to fainter, sub-Eddington bursts, as witnessed by the
detection of B3 and B13 (Table~\ref{table:bprop} \&
Figure~\ref{fig:bspec}).
All the GBM bursts presented herein are much shorter than the segment
of the long burst detected by FREGATE. This, together with the
simulations described below, indicates that all bursts B1--B15 are
normal duration bursts, even after bandpass corrections are applied.

Due to its spectral evolution, a given burst can produce very
different light curves when observed with instruments with different
bandpasses \citep[see, e.g.,][]{Chelovekov06}. This is of particular
relevance when studying thermonuclear bursts with a hard X-ray
detector, as many of the burst photons are emitted outside the energy
range to which the instrument is sensitive to. Whenever a burst is
clearly detected, its thermal spectrum is usually well constrained
using the hard X-ray band and the bolometric flux can therefore be
measured, under the assumption that color and effective temperatures
are equal.

However, the actual burst detection relies on having sufficient burst
photons emitted in the instrument band to rise significantly above the
background rate. In practice, this implies that instruments like GBM
\citep[sensitive to photon energies above 8~keV;][]{Meegan09}, FREGATE
\citep[sensitive above 6~keV;][]{Atteia03}, BAT or IBIS-ISGRI
\citep[sensitive above 15~keV;][respectively]{Barthelmy05,Lebrun03}
are able to detect only the hottest ($kT_\mathrm{bb}
\gtrsim$1.5--2~keV) and relatively long-lived phases of bursts (see
Figure~\ref{fig:iceberg}).
Thus in general the radiated energy and duration measured in the hard
X-ray band ($E_\mathrm{obs}$ and $t_\mathrm{obs}$, respectively) are
lower limits to the total burst fluence and duration ($E_\mathrm{b}$
and $t_\mathrm{b}$). This was illustrated by the long burst from
XTE~J1701--407 detected with both the XRT (sensitive in the
0.5--10~keV band) and BAT instruments onboard {\it Swift}
\citep{Linares09b}. In that case $E_\mathrm{b}$ and $t_\mathrm{b}$
exceeded those measured in the BAT (15--50~keV) band by a factor of
$E_\mathrm{b}/E_\mathrm{obs} \simeq$3.9 and
$t_\mathrm{b}/t_\mathrm{obs} \simeq$20, respectively
\citep{Linares09b}.

In order to quantify this combined bandpass and background effect for
the GBM bursts from 4U~0614+09 and to estimate the total radiated
energy and duration, we simulated GBM light curves using two of the
bursts reported by \citepalias{Kuulkers10}: the long PRE burst shown
in Figure~\ref{fig:iceberg} and a normal duration burst. We created a
response matrix for one GBM NaI detector (n0, 14$^\circ$ off-axis;
CTIME mode) and used the spectral parameters from both bursts (with a
10~s time resolution) to simulate net light curves within Xspec
\citep[v. 12.7.0;][]{Arnaud96}. We calculated count rates in the energy
range 12--25~keV (CTIME channel 1), the band used in our systematic
search (Sec.~\ref{sec:search}). We then added the net burst counts to a
real background light curve which includes contributions
from particle background and X-ray sources in the field of view, both
variable typically on timescales $\gtrsim$100~s. Finally, we measured
the energy (from the integrated fluence, using the 3.2~kpc distance)
and duration of each burst in the sub-interval detected by GBM.

\begin{figure*}[ht]
\centering
  \resizebox{1.5\columnwidth}{!}{\rotatebox{-90}{\includegraphics[]{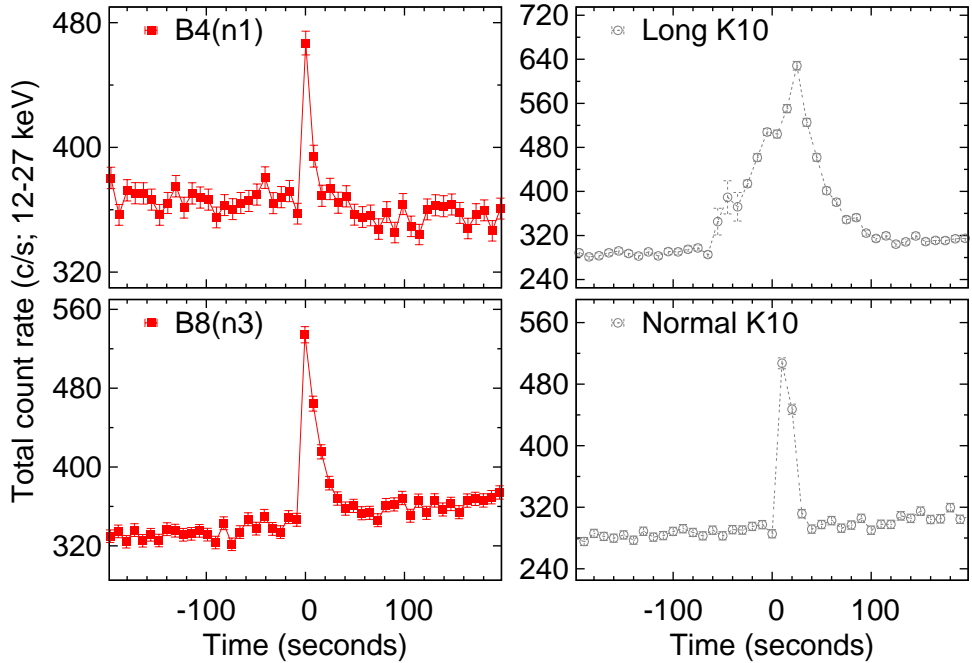}}}
  \caption{
{\it Right panels:} Simulated GBM light curves in the 12-25~keV band
  at 10~s time resolution for a pointing offset of 14$^\circ$, using
  the spectral evolution measured by HETE-FREGATE during a long ({\it
  top}) and normal duration ({\it bottom}) burst from 4U~0614+09
  \citepalias[both taken from][]{Kuulkers10}.
{\it Left panels:} Light curves for two of the GBM bursts from
  4U~0614+09 reported herein, in the 12--25~keV band with 8~s time
  resolution and including background rate. In both cases the pointing
  offset was 14$^\circ$; detector number is indicated between
  parenthesis next to the burst ID (Tables~\ref{table:bursts} and
  \ref{table:bprop}).
}
    \label{fig:simulate}
\end{figure*}

The resulting simulated light curves are shown in
Figure~\ref{fig:simulate} (right panels). The long burst seen in the
GBM band would present a characteristic slow rise
($t_\mathrm{rise}$$\simeq$65~s) and last for a total of
about 140~s. The simulated normal duration burst, on the other hand,
has a rise time (6.4~s), energy (3.6$\times$10$^{39}$~erg) and
duration (30~s) similar to all GBM bursts presented herein
(Table~\ref{table:bprop}). This is shown in Figure~\ref{fig:simulate}
by comparing two of the GBM bursts (left panels; B4 \& B8) with the
simulated long and normal burst light curves (right panels). We
conclude that all bursts detected from 4U~0614+09 during the first
year of our GBM X-ray burst search are normal duration bursts, yet as
discussed below they are more energetic than most normal duration
bursts.

Table~\ref{table:bolo} presents the duration and energy of the long
and normal duration bursts, as measured with three different
instruments and bandpasses: GBM, FREGATE and PCA. The GBM values are
estimated from the simulations explained above, while FREGATE and PCA
values are taken from \citetalias{Kuulkers10}. The bolometric
correction factors for energy and duration are given for each case.
FREGATE was able to detect a slightly longer portion of the
bursts. Although FREGATE also consisted of NaI scintillation detectors
with effective area similar to that of the GBM detectors, its lower
energy threshold (6~keV) and smaller field of view (which reduces
X-ray background) made it sensitive to a larger fraction of the
bursts.

The bandpass used has a clear effect on the measured burst energy,
with $E_\mathrm{b}/E_\mathrm{obs}$ in the range 1.3--1.7 \citep[1.3--4
if we include the XRT/BAT long burst from][]{Linares09b}. We note that
the same is true for the commonly used characteristic burst timescale
$E_\mathrm{b}/L_\mathrm{peak}$. The impact on the measured burst
duration is much more drastic, as $t_\mathrm{b}/t_\mathrm{obs}$ can be
anywhere between 1.5 and 55. This is because cooling tails can
extend for a long time \citep[e.g.;][]{Zand09} and are detected only
with soft X-ray detectors, yet most of the energy is radiated during
the hottest, most luminous phases of bursts which are captured to a
greater extent by hard X-ray detectors. Thus in general burst energy
constitutes a more robust (less band-dependent) observable than burst
duration.
As normal duration bursts from 4U~0614+09 have not been observed with
pointed soft X-ray detectors, we cannot find the bolometric correction
factors for the case of normal duration bursts from 4U~0614+09
observed with GBM. Based on our GBM simulation and the FREGATE normal
duration burst, we estimate that in this case
$E_\mathrm{b}/E_\mathrm{obs} \simeq$2 (Table~\ref{table:bolo}), i.e.,
normal bursts are about twice as energetic as seen in the GBM band.

\begin{figure*}[ht]
\centering
  \resizebox{1.6\columnwidth}{!}{\rotatebox{0}{\includegraphics[]{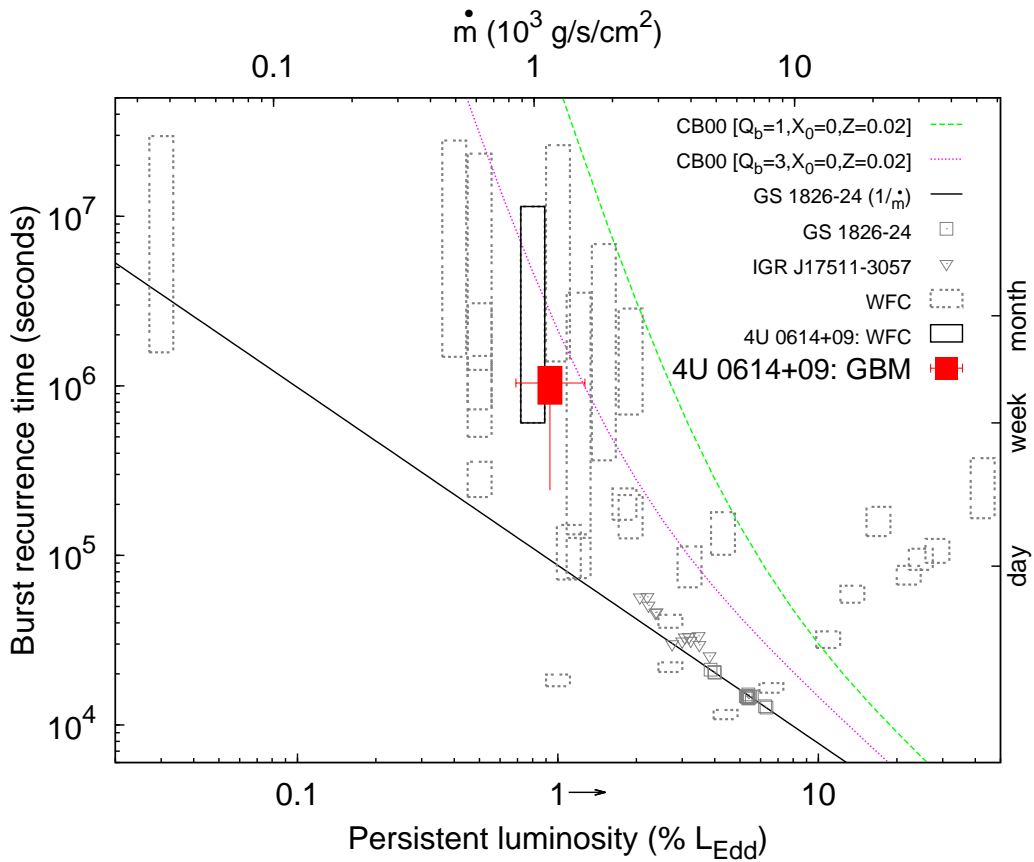}}}
  \caption{Burst recurrence time $t_\mathrm{rec}$ vs. persistent
  luminosity and inferred $\dot{m}$ for 4U~0614+09 and a sample of
  bursters. The red filled rectangle shows the GBM measurement
  (1$\sigma$ confidence interval) of the average $t_\mathrm{rec}$ in
  4U~0614+09. The red vertical error bar corresponds to the shortest
  wait time between bursts that we find, 2.8~d, and the horizontal
  error bars show the range in $L_\mathrm{pers}$ and $\dot{m}$. Open
  (gray dashed) rectangles show the 1$\sigma$ confidence intervals
  obtained from 6~yr of WFC data, taken from \citet[][only systems
  below 50\%~$L_\mathrm{Edd}$ are shown]{intZand07}. The solid open
  black rectangle shows the WFC limits on $t_\mathrm{rec}$ for
  4U~0614+09. Open squares and triangles show $t_\mathrm{rec}$
  measurements in GS~1826-24 \citep{Galloway04} and IGR~J17511-3057
  \citep{Falanga11}, respectively. The solid black line shows the
  $t_\mathrm{rec} \propto 1/\dot{m}$ relation found by
  \citet{Galloway04} for GS~1826-24. Green dashed and purple dotted
  lines show the $t_\mathrm{rec} (\dot{m})$ relation predicted by He
  ignition models \citep[][for pure He accretion and solar
  metalicity]{CB00} for a base heat flux of 1~MeV/nucleon and
  3~MeV/nucleon, respectively. The arrow below the bottom axis
  shows the change on the Eddington fraction if a value of
  $L_\mathrm{Edd}$=3.8$\times$10$^{38}$~erg~s$^{-1}$ is used
  (Sec.~\ref{sec:pers}).}
    \label{fig:trecmdot}
\end{figure*}

\subsection{4U~0614+09 and low-$\dot{m}$ bursts}
\label{sec:beef}

In Figure~\ref{fig:trecmdot} we compare the $t_\mathrm{rec}$ measured
in 4U~0614+09, when $\dot{m}$ was between 0.6\% and 1.3\%, to the
constraints available for other low-$\dot{m}$ bursters and a few
selected systems. The $t_\mathrm{rec}$-$\dot{m}$ relations for two
``well-behaved'' bursters accreting at a few percent of
$\dot{m}_\mathrm{Edd}$ are shown for comparison
\citep{Galloway04,Falanga11}. In those cases $t_\mathrm{rec}$ is
shorter than a day and thus much easier to measure. For comparison, He
ignition curves for pure He accretion are also shown, for two
different values of the heat flux from the crust
\citep[$Q_\mathrm{b}$=1~MeV/nucleon and $Q_\mathrm{b}$=3~MeV/nucleon;
see][for details]{CB00}. The large change in predicted
$t_\mathrm{rec}$ illustrates the sensitivity of low-$\dot{m}$ bursts
to the thermal properties of the crust.
These models, however, predict burst energies much higher than the
observed ones. Detailed modeling of the burst properties is beyond the
scope of this work, and we refer the reader to \citetalias{Kuulkers10}
for further discussion.

\begin{figure*}[ht!!]
\centering
  \resizebox{1.6\columnwidth}{!}{\rotatebox{0}{\includegraphics[]{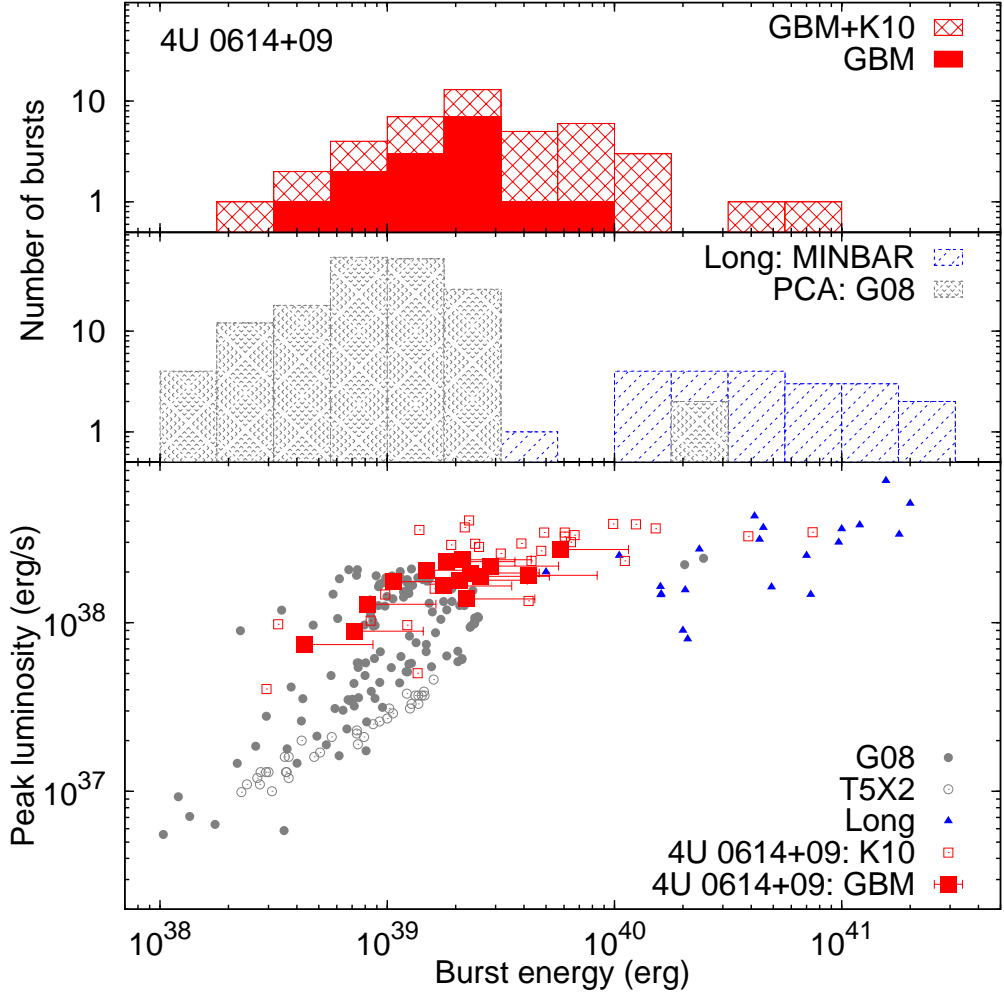}}}
  \caption{{\it Bottom:} Burst peak luminosity vs. energy, for the GBM
  bursts from 4U~0614+09 presented herein (filled red squares; error
  bars show estimated bolometric correction factor, see
  Sec.~\ref{sec:band}), for all known bursts from 4U~0614+09 including
  those reported in \citetalias{Kuulkers10} (open red squares), for
  all bursts from the {\it RXTE} catalog \citep[filled gray
  circles,][]{Galloway08}, for the 11~Hz pulsar in the globular
  cluster Terzan~5 \citep[open gray circles,][]{Linares12} and for the
  long/intermediate bursts from the MINBAR catalog (filled blue
  triangles). {\it Middle and top panels:} Histograms showing the
  number of bursts per energy bin for the different known samples, as
  indicated. 4U~0614+09 bridges the gap between the ``normal'' and
  ``long/intermediate'' burst populations (Sec~\ref{sec:beef}).}
    \label{fig:hist}
\end{figure*}

It is interesting to compare the distribution of burst energies from
4U~0614+09 with that of the normal duration and long/intermediate
burst populations (see Figure~\ref{fig:hist}). As discussed in
Section~\ref{sec:band}, burst energy is less sensitive to bandpass
effects than burst duration, and may constitute a more robust
observable for burst classification. 4U~0614+09 has on average larger
$E_\mathrm{b}$ than normal duration bursts, yet smaller $E_\mathrm{b}$
than long/intermediate duration bursts. The energies of the two long
bursts from 4U~0614+09 \citepalias{Kuulkers10}, however, are
consistent with the rest of the long burst population. Even after
bandpass effects are corrected for, 4U~0614+09 bridges the gap between
normal duration and long/intermediate bursts (Fig.~\ref{fig:hist}).
This could be due to a fuel composition different than most
ultracompact LMXBs that would give rise to atypical bursts, if as it
has been suggested the donor star has unusual abundances \citep[H and
He defficient;][see also Sec.~\ref{sec:intro}]{Werner06}.

Alternatively, the previously observed bimodal distribution of burst
durations could be due to a selection effect. Prior to our GBM
campaign, observations of low-$\dot{m}$ bursts were mostly based on
serendipitous detections with hard X-ray wide-field instruments. If
such serendipitous and so far relatively scarce detections are biased
towards the longest and most energetic events, a population of
infrequent, low-$\dot{m}$ bursts with energies and durations between
normal and long bursts could have gone unnoticed. Our GBM all-sky
search (Sec.~\ref{sec:search}) is the first systematic survey of
low-$\dot{m}$ bursts and is less biased towards long and energetic
events.
A continuous distribution in $E_\mathrm{b}$ (between 10$^{39}$~erg and
10$^{41}$~erg) is actually not surprising from the standpoint of He
ignition models \citep{CB00}. This is, as far as we know, the first
time that such continuous distribution is observed.
Extending this study to similar bursters will determine if the
distribution of burst energies that we find (Figure~\ref{fig:hist}) is
unique to 4U~0614+09 or a common feature of low-$\dot{m}$ bursters.

The average burst energy in our sample, after the bolometric
correction factor of $\simeq$2 is applied (Sec.~\ref{sec:band}), is
5$\times$10$^{39}$~erg. For the measured $t_\mathrm{rec}$ (12~d) and
$L_\mathrm{pers}$ (2.3$\times$10$^{36}$~erg~s$^{-1}$), this
corresponds to an accretion-to-burst fluence ratio
$\alpha$~$\simeq$~560, while the accreted column depth at the average
inferred $\dot{m}$ (1.1$\times$10$^3$~g~cm$^{-2}$~s$^{-1}$) is
$y$~$\simeq$~1$\times$10$^9$~g~cm$^{-2}$, similar to the values
obtained from light curve fitting
\citepalias[2$\times$10$^9$~g~cm$^{-2}$ for the normal duration burst
in][]{Kuulkers10}.
However, complete ignition of such a thick layer would give rise to
more energetic bursts than observed, assuming the energy release
expected for pure He burning \citepalias[as discussed
by][]{Kuulkers10}.
For the shortest $t_\mathrm{wait}$ burst (B13), using the average
$L_\mathrm{pers}$ and $\dot{m}$ during January 2011
(1.8$\times$10$^{36}$~erg~s$^{-1}$ and
0.9$\times$10$^3$~g~cm$^{-2}$~s$^{-1}$, respectively) yields a
similarly high value of $\alpha$~$\simeq$506. This indicates that
even the shortest $t_\mathrm{rec}$ bursts in 4U~0614+09 are
sub-energetic.
Thus after our improved $t_\mathrm{rec}$ measurements and considering
bolometric/bandpass corrections to $E_\mathrm{b}$, the question of why
are most bursts in 4U~0614+09 sub-energetic remains open.
We also note that the shortest $t_\mathrm{wait}$ burst (B13) has
the lowest $E_\mathrm{b}$ in our sample, as expected if the mass
accumulated before ignition is also lowest.
We found the period with the shortest $t_\mathrm{rec}$ near the
minimum in $\dot{m}$ (Sec.~\ref{sec:results}). This behavior is
opposite to what ignition models predict and it might be due to
different rates of stable H (CNO) burning operating at slightly
different $\dot{m}$ (although the apparent lack of H in the fuel makes
this scenario unlikely). Alternatively, it could reflect changes in
the $L_\mathrm{pers}$-$\dot{m}$ relation (i.e., in the radiative
efficiency of the accretion flow) unaccounted for. Continued
monitoring may determine whether this is a weeks-long fluctuation in
the observed burst rate or it constitutes a systematic behavior of
4U~0614+09.

As can be seen in Figure~\ref{fig:trecmdot}, our GBM campaign in its
first year has improved the accuracy of $t_\mathrm{rec}$
measurements at low $\dot{m}$, placing new constraints on models of
low-$\dot{m}$ ignition.
Most estimates available to date were based on {\it BeppoSAX}-WFC
observations and provided mostly order-of-magnitude constraints on
$t_\mathrm{rec}$ (Fig.~\ref{fig:trecmdot}).
Extending our GBM measurements to other low-$\dot{m}$ bursters and
comparing them to theoretical models, considering potential changes in
fuel composition across bursters, can make precise measurements of
crustal heat flux possible.

\section{Summary}
\label{sec:summary}

\begin{itemize}[leftmargin=0.2cm]

\item We have presented 15 normal duration thermonuclear bursts from
4U~0614+09 observed with {\it Fermi}-GBM between March 2010 and March
2011, when the source was accreting near 1\% of the Eddington rate. We
measured a burst recurrence time of 12$\pm$3~d and find the closest
burst pair seen from 4U~0614+09 to date, 2.8~d apart.

\item We find no long/intermediate bursts in one year of data, and set
a lower limit to their recurrence time of 62~d. The 2001--2002
``calm'' (low long-term variability) period in the persistent soft
X-ray emission during which long/intermediate bursts were detected has
not recurred.

\item We quantify the bandpass effect on the observed burst duration
and energy. We find that the total burst duration can be 1.5--55
times longer than that measured in the hard X-ray band, whereas this
bolometric correction factor is 1.3--1.7 for burst energy (which is
therefore a more robust, band-independent observable).

\item The burst energies in 4U~0614+09, between 8$\times$10$^{38}$~erg
and 1$\times$10$^{40}$~erg after our bolometric correction is applied,
overlap between the energies of the normal and long burst
populations. This continuous distribution in burst energy provides new
observational evidence that long/intermediate bursts are an extreme
case of low-$\dot{m}$ bursts.
We suggest that the apparent bimodal distribution of
durations that defined normal and long/intermediate bursts could be
due to a selection effect if only the longest and most energetic
low-$\dot{m}$ bursts could be detected to date.

\end{itemize}

\textbf{Acknowledgments:}
Quick-look ASM results were provided by the RXTE/ASM team. Swift/BAT
transient monitor results were provided by the Swift/BAT team. This
paper utilizes preliminary analysis results from the Multi-INstrument
Burst ARchive (MINBAR;
http://users.monash.edu.au/$\sim$dgallow/minbar). Data from previous
bursts were kindly provided by E. Kuulkers. We thank A. Cumming for
providing ignition models and for stimulating discussions and
V. Chaplin for clarifications on detector geometries during some of
the bursts. We also thank the anonymous referee for constructive
comments. ML is grateful to the International Space Science Institute
in Bern, where part of this work was completed, and acknowledges
support from the NWO Rubicon fellowship. This research was partly
funded by NASA's Fermi Guest Investigation program under grant
NNX11AO19G.

\bibliographystyle{apj} 


\end{document}